\documentstyle[]{article}
\begin{document}
\begin{center}
{\Large \bf  Landau theory of
\( 180^{o} \) domain walls
in \( BaTiO_{3} \) type ferroelectric particles: microcomposite
materials}
\end{center}
\vskip4cm
\begin{center}
{\large \bf O. Hudak  \\
\vskip2cm
Department of Theoretical Physics, Faculty of Mathematics, Physics and Informatics, Comenius University, Mlynska dolina F2, 845 01 Bratislava, Slovakia
\\
}
\end{center}
\begin{center}
{\large \bf
\vskip2cm
Matej Hudak
\vskip1cm
Stierova 23, SK - 040 23 Kosice, Slovakia
}
\end{center}

\newpage
\section*{Abstract}
The Landau theory of
\( 180^{o} \) domain walls
in \( BaTiO_{3} \) type ferroelectric particles is
presented . Results of exact description of domain walls in
bulk enabled us to formulate variational approach to theory of
domain walls in corresponding small particles.
The depolarization field effects and the space-charge layers are
taken into account in the samples of the cube form.
It was found that at low temperatures well known hyperbolic tangent
wall profile is a good approximation for
description of domain walls  in particles.
Near the transition temperature
it is more and more appropriate to speak about two walls separating
ferroelectric, paraelectric and ferroelectric domains correspondingly
as a result of splitting of
a single ferroelectric wall into two subwalls in
small \( BaTiO_{3} \) particles.
Domain  wall energy density, 
average interwall distance and change of the dielectric response of thick
walls in small ferroelectric particles in microcomposites is qualitatively found. Our results qualitatively describe observed dependencies better than those theories which exist up to date. In temperature region near transition from the ferroelectric to paraelectric phase in micrcomposites our quantitative results for response, for Curie-like transition temperature and other parameters may be verified.

\newpage
\section{Introduction}
The transition
temperature of the ferroelectric phase - paraelectric
phase phase transition decreases
with decreasing particle diameter in $\rm BaTiO_{3}$ \cite{uchino},
$\rm PbTiO_{3}$ \cite{ishikawa} and in KDP
\cite{jona} microcomposites. It was found experimentaly that
there is no phase transition bellow a critical size of
a particle in the microcomposite. It was observed also that a higher phase transition temperature in smaller particles sometimes
occurs. In thin films of $\rm KNbO_3$ \cite{scott}
and TGS \cite{hadi,batra} both increase and decrease of the transition temperature
were observed with decreasing particle size.
In $\rm PbTiO_3$ microcomposites \cite{lee}
the dielectric constant decreases with decreasing particle
size, probably due to
the gradual creation of a multidomain state in larger
particles.
An unusual peak of dielectric constant versus particle size was
observed in $\rm BaTiO_3$ \cite{arlt}.

There are two main theoretical approaches to explanation of the size
effects in the ferroelectric - dielectric microcomposites of the above type.
The significance  of depolarization field that tries to break up the ferroelectric particle into domains with different polarizations. The multidomain state then exhibits higher
permittivity, but its creation becomes difficult for small enough
particles \cite{shih}. Another possibility is that a surface layer with different transition temperature than that in the bulk is phenomenologically considered without
the depolarization field explicitly  taken into account. Then
a shift of the phase transition temperature in the particle is present as well as the inhomogeneous distribution of the polarization \cite{zhong}. In \cite{zhong} the static properties of such particles is studied by numerical calculations only. The ferroelectric phase
transition in small spherical particles was studied in \cite{RH}. Size effects were studied
analyticaly  as concerning temperature effects and
polarization profile. Our model in \cite{RH} did not take into
account depolarization  fields, assuming full compensation  of surface charges. The dynamic
susceptibility deviates from Debye-like behaviour then and exhibits a broadening at higher frequencies. The Curie-Weiss law is exhibited by the static
susceptibility, there exists a divergency at the
point of the size-driven transition.
Dielectric response of microcomposite
ferroelectrics was studied in \cite{HRP}.
A two-phase composite of ferroelectric and dielectric
materials was described, and studied by different
effective medium theories (isolated particles,
Maxwell-Garnet theory, the effective medium
approximation and the Bergman representation of
the effective dielectric function). There are
present modes due to geometric resonances which
lead to new low-frequency absorption peaks near
the percolation threshold of the ferroelectric
material. Those peaks are always located higher
than the peak of the transverse polar mode of the
bulk component. The soft mode with renormalized
mode strength but unchanged frequency exists only
in the composite with percolated ferroelectric
clusters. Below the percolation threshold the
finite clusters of the ferroelectric material
soften only to some degree, there is no soft mode.
Results of exact description of domain walls in
bulk have been found and were preliminary published in \cite{COMATECH99}. They enabled us to formulate variational approach to the theory of domain walls in corresponding small particles.

In this paper we study a Landau theory of
\( 180^{o} \) domain walls
in \( BaTiO_{3} \) type ferroelectric particles.
The depolarization field effects and the space-charge layers are
taken into account in the samples of the cube form. 
We have found that at low temperatures well known hyperbolic tangent
wall profile is a good approximation for
description of domain walls  in particles.
Nearby the first-order
transition temperature a free energy contribution
due to paraelectric domain state 
within a wall center becomes important leading to
unusually large wall thickness.
Approaching the transition temperature
it is more and more appropriate to speak about two walls separating
ferroelectric domains as a result of splitting
a single ferroelectric wall into two walls with a paraelectric phase between them.
Consequently the primary
ferroelectric domain wall width becomes very large
which influences also size dependence and other characteristics
of the ferroelectric transition to the paraelectric phase
in small \( BaTiO_{3} \) particles.
Domain  wall energy density,
average interwall distance and change of the dielectric response of thick
walls in small ferroelectric particles is qualitatively described and quantitatively approximately found in
this temperature region. 

\section{Domain walls configurations in a \( BaTiO_{3} \) particle}

The free energy density F expansion
into polar mode amplitude {\bf P} following \cite{SSA} has the form
\begin{equation}
\label{1}
F= \frac{\alpha}{2}{\bf P}^{2} + \frac{\beta}{4}{\bf P}^{4} +
\frac{\sigma}{6}{\bf P}^{6} + C \mid {\bf \nabla P} \mid^{2}.
\end{equation}
where the coefficient is given by \( \alpha = a_{0} (T - T_{0 \infty}), \) here \(
a_{0} \) is a constant, \( T_{0 \infty} \) is the Curie-Weiss temperature for the bulk phase transition.

Numerical values of the coefficients of the Landau free energy
expansion (\ref{1}) of \( BaTiO_{3} \) are given in \cite{SSA}.

Assuming the  domain wall localized at x=0,  the
polarization {\bf P} oriented in the z-direction and that this polarisation
changes in the x direction, P(x), the domain-wall energy \( \gamma
\) is defined by
\begin{equation}
\label{2}
\gamma=\int^{+\infty}_{-\infty} \triangle F(P(x)) dx,
\end{equation}
where \( \triangle F(P(x))=F(P(x))-F(P) \) is the difference of free
energy densities between the state with a domain-wall configuration P(x)
and the equilibrium state with a homogeneous polarization P.

The single domain-wall configuration P(x) in \( BaTiO_{3} \) was in
\cite{SSA} described approximately by the following well-known profile
\begin{equation}
\label{3}
P(x)= P.tanh(\frac{x}{\xi}),
\end{equation}
with P and \( \xi \) as variational parameters.
The domain-wall configuration P(x) in a \( BaTiO_{3} \) particle has a wall
energy density which may be approximately calculated as in \cite{SSA} by
\begin{equation}
\label{4}
\gamma=\int^{+D}_{-D} \triangle F(P(x)) dx,
\end{equation}
with 2D a distance between two neighbouring walls. It is a variational parameter.
After calculating the wall energy density \( \gamma \) one can then easily
incorporate space-charge layer effects, characterised by a parameter t,
in the same way is in \cite{SSA}, too.

\section{Exact solution for wall configurations}

Let us transform the free energy part (\ref{1}) describing
local potential of the sixth order in the polarization P to a more
transparent form
\begin{equation}
\label{11}
V(P)= b[\frac{P^{6}}{6}-(a^{2}+c^{2})\frac{P^{4}}{4}+a^{2}c^{2}
\frac{P^{2}}{2}].
\end{equation}
The coefficients a, b and c above and the free energy density
expansion coefficients \( \alpha, \beta, \gamma
\) are connected by relations:
\begin{equation}
\label{11'}
a^{2}=\frac{1}{2}[\frac{\beta}{\sigma}+\sqrt{\frac{\beta}{\sigma}-
\frac{4\alpha}{\sigma}}],
\end{equation}
\[ c^{2}=\frac{1}{2}[\frac{\beta}{\sigma}-\sqrt{\frac{\beta}{\sigma}-
\frac{4\alpha}{\sigma}}], \]
\[ b=\sigma. \]

To discuss properties of the local effective
potential (\ref{11}) it is convenient to introduce a
dimensionless parameter \( \rho \) 
\begin{equation}
\label{12}
\rho = (\frac{c}{a})^{2},
\end{equation}
which we will assume to be a nonegative quantity. The case of
negative values leads to a two well potential of the sixth order,
which is similar to the well-known double-well potentials.

The local potential V(P) has two absolute minima at \( \pm a \) and
there is a local maximum at P=0 for \( \rho=0. \)
The potential V(P) has two absolute minima \( \pm a \) and
there is a local minimum at P=0, and at \( \pm c \) there are two
local maxima for \( 0< \rho < \frac{1}{3}. \)
The potential V(P) has three absolute minima at \( \pm a \) and
at P=0, in \( \pm c \) there are two local maxima for \( \rho
= \frac{1}{3}. \)
The potential V(P) has only single absolute minimum at P=0, at \(
\pm a \) there two local minima and
in \( \pm c \) there are two local maxima for \(
\frac{1}{3} < \rho <1. \)
For \( \rho >1 \) the situation just described repeats with the
reversed role of the a and c constants.

The Lagrange-Euler equation for configurations of polarisation P which extremize the free energy
(\ref{1}) has the form
\begin{equation}
\label{13}
2C \frac{dP}{dx}= b P.(P^{2}-a^{2}).(P^{2}-c^{2}).
\end{equation}
We assume that \( a > c, \) the equilibrium state configurations  then realizes
at points \( \pm a \) (ferroelectric phase) or at P=0 (paraelectric phase).

Severals years ago \cite{OH} we succeeded  to find exact solutions of the equation
(\ref{13}). Similar solutions in different forms were obtained
also elsewhere, \cite{F} and \cite{CC}. For \( 0 \leq \rho < \frac{1}{3} \)
the walls are described in the form
\begin{equation}
\label{14}
P(x)= \pm a \frac{\sqrt{q}tanh(\frac{x}{\xi}+R)}{\sqrt{1+q-tanh^{2}(\frac{x}{\xi}+R))}}
\end{equation}
where R is an arbitrary constant, describing the position of the
center of the wall, and where we define \( q =
\frac{1-3\rho}{3}. \)
These walls interconnect two ferroelectric domains with opposite
polarization.

The parameter \( \xi \) describes the rate of change of the
polarization at the wall configuration and is given by
\begin{equation}
\label{16}
\xi
=\frac{1}{\sqrt{\frac{1+q}{3}}}\frac{1}{\sqrt{\frac{ba^{4}}{2C}}}.
\end{equation}

\section{\( 180^{o} \) domain wall configurations}

The exact solution (\ref{14}) describes a profile of a domain wall. It is convenient to rewrite the form (\ref{14}) into a more transparent form
\begin{equation}
\label{17}
P(x)= P_{0}(x) . tanh(\frac{x}{\xi}+R),
\end{equation}
where the first factor in (\ref{17}) we call a  "modulated amplitude" \( P_{0}(x) \) of the wall configuration tanh. It is given by
\begin{equation}
\label{18}
P_{0}(x)=  \pm a \frac{\sqrt{q}}{\sqrt{1+q-tanh^{2}(\frac{x}{\xi}+R))}}.
\end{equation}

While at larger distances from the center of the wall, localized at \( x=0 \) for R=0,
the "modulated amplitude" takes the equilibrium configuration values
\begin{equation}
\label{19}
P_{0}(\pm \infty)=  \pm a.
\end{equation}

at the center of the wall this amplitude decreases its value to
\begin{equation}
\label{20}
P_{0}(x=0)=  \pm a \frac{\sqrt{q}}{\sqrt{1+q}}= P_{0}(\pm \infty). \sqrt{\frac{1-3\rho}{3-3\rho}}.
\end{equation}

It is clear from (\ref{20}) that the nearer we are to the phase transition
the nearer the constant \( \rho \) to the value \( \frac{1}{3}, \) and
consequently the smaller "modulated amplitude" of the wall at the center. This decrease
reflects the increasing role of the paraelectric phase region in the center of the wall.

The wall thus consists of two subwalls connecting domains with nonzero
polarization with a region with almost zero polarization. At temperatures
near the phase transition temperature
the distance \( d_{w} \) between the two subwalls (walls
separating the ferroelectric domains and the paraelectric phase domain in the center)
becomes larger. One can find that at this temperature
region we have the distance \( d_{w} \) given by
\begin{equation}
\label{21}
d_{w}=-\frac{\xi}{2} ln(\frac{1-\sqrt{3\rho}}{1+\sqrt{3\rho}}),
\end{equation}
which describes logarithmic  divergence of the distance \( d_{w} \), because the nearer we are to the phase transition the nearer the constant \( \rho \) to the value \( \frac{1}{3}. \)  

On the other hand the lower temperature the larger the "modulated amplitude" \( P_{0}(x=0),\)
approaching the same value as the amplitude at domains \( P_{0}(\pm \infty). \)
One can find that in this temperature region
\begin{equation}
\label{22}
d_{w}= \xi \sqrt{3\rho},
\end{equation}
which is a finite quantity.

One can easily estimate temperature region in which the condition \( 0 < \rho  <
\frac{1}{3} \) is satisfied and
thus in which very thin walls may exist the distance \( d_{w} \) of which logarithmicaly  divergences. Using parameters for \( BaTiO_{3} \)
from \cite{SSA} we found that
the mentioned inequalities are equivalent with temperature region
\( 112.24^{o} C < T < 121.48^{o} C. \) In this region thus the ferroelectric domain wall has in the center the large paraelectric region and consists of two walls /subwalls/ connecting this region with the corresponding ferroelectric region.

From this result
it follows that at \( 25^{o} C \) the "tanh" wall profile, used in \cite{SSA}
for approximate calculations, is an adequate approximation.
At this lower temperatures this material is characterized by an effective
potential with only two absolute minima. Near phase transition from the ferroelectric phase to the paraelectric phase in the temperature region
\( 112.24^{o} C < T < 121.48^{o} C \) the "tanh" wall profile used for approximate calculations is an inadequate approximation. Our exact description of the domain wall is more appropriate in this region.

\section{The dielectric response of a ferroelectric particle}

The dielectric response of a ferroelectric particle at temperature region near
the transition temperature, where the wall width behaves according
to (\ref{22}), may be estimated in the following way. The effect of the
depolarization field and of the space-charge layer may be taken into account
using the same variational approach as in \cite{SSA}. However, as the wall
thickness parameter the domain subwall distance as a domain wall width \( d_{w} \) instead of
\( \xi \) from \cite{SSA} is taken.

This approach leads to the following wall energy density 
\( \gamma \)
\begin{equation}
\label{22.1}
\gamma \approx \frac{8 c P_{\infty}^{2}}{3 d_{w}}.
\end{equation}

The equilibrium interwall distance 2D between two ferroelectric walls in a particle of a cube form with the side width L is found to be given by D
\begin{equation}
\label{22.2}
D = \sqrt{\frac{8 c}{5.1 d_{w}t}}L.
\end{equation}
where t characterizes changes of the polarisation near the surface. 
We see that increasing temperature to the transition temperature
the domain wall width measured by the subwall distance \( d_{w} \) increases while the interwall
distance D decreases. Remaining structure consists of strips
of ferroelectric phase domains separated by wide regions
(domains) of the paraelectric phase.

If both characteristic lengths, the interwall distance characteristic quantity D and the subwall distance characteristic quantity \( d_{w} \)  become comparable, \( D
\approx d_{w}, \) then the ferroelectric domains become very
narrow and almost vanish with respect to usual situation in which ferroelectric domain wall width is very small with respect the ferroelectric domain width. The corresponding temperature is
slightly below the transition temperature at which \( \rho = \frac{1}{3}. \)

In the regime in which \( \xi << D, \) one finds the Curie-Weiss law with renormalised transition temperature
\begin{equation}
\label{23}
\alpha^{'}=\alpha + 3.4 \frac{Dt}{L^{2}}+[-\frac{2d_{w} \alpha}{D}+ \frac{8c}{3 d_{w}}].
\end{equation}
From this expression one finds that the temperature coefficient a is multiplied
by \( 1-\frac{2d_{w}}{D} \) giving \( a^{'}. \)

The critical
Curie-Weiss temperature becomes
\begin{equation}
\label{24}
T_{0} = T_{0\infty}-\frac{1}{a'}[3.4 \frac{Dt}{L^{2}}+ \frac{8c}{3 d_{w}}].
\end{equation}

If the wall thickness (\ref{21}) increases increasing temperature there will
be much stronger renormalization of the temperature coefficient to zero
approximately for those wall configurations at which the the interwall
distance 2D becomes comparable with the width parameter \( d_{w}. \)
From (\ref{24}) it follows that simultaneously the Curie-Weiss temperature
reduces to zero quickly, the ferroelectric phase vanishes even before the wall
thickness becomes comparable with the interwall distance characteristic length
D. It is also clear that the existence of the ferroelectric phase becomes more sensitive to
the length L of the crystal due to decreased value of the temperature
coefficient \( a^{'} \) and due to the domain wall
width \( d_{w} \) dependence of the interwall
distance 2D (\ref{22.2}).

For thin walls $(d_{w} < \frac{D}{8})$ the transition
temperature dependence coefficient on the particle size L
is larger for the domain walls as described in our
paper than for those described approximately
by simple $tanh$ dependence used in \cite{SSA} to
qualitatively describe transition temperature
dependence on L.
Experimentally found transition temperature
dependence on L is steeper than that predicted by
the simple tanh wall model \cite{SSA}.
Our exact description od domain walls
for the bulk free energy potential in \( BaTiO_{3}
\) describing the first order phase transition which
is near the second order phase transition and used
for the description of the particle domain
structure as in \cite{SSA}
is more appropriate for real crystals \( BaTiO_{3}
\). Note that for
thick walls $(d_{w} > \frac{D}{8})$ the transition
temperature dependence  coefficient on the particle size L
is smaller for the domain walls as described in our
paper than for those described approximately
by simple $tanh$ dependence.
Thus mechanism of the increasing  domain wall width
due to splitting of the domain wall into two walls
(sequence ferroelectric - paraelectric -
ferroelectric domains) may explain much more abrupt change
of the Curie-Weiss temperature al low L values unexplained by the spontaneous
strain mechanism in \cite{SSA}.

\section*{Acknowledgement}

The author wishes to express his sincere thanks to V.Dvorak, J.Petzelt,
I.Rychetsky, J.Holakovsky and M.Glogarova  for their
discussions about microcomposite materials.
This paper was supported by the grant VEGA 1/0250/03, Faculty of Mathematics, 
Physics and Informatics, Comenius University, Bratislava.

\end{document}